\documentclass[draft]{agujournal2019}
\usepackage{color}
\usepackage{url} 
\DeclareUrlCommand\url{\color{blue}}
\usepackage{lineno}
 \usepackage{graphicx}
 \usepackage{amsmath,amsfonts}
\usepackage{algorithm}
\usepackage{algpseudocode}
\usepackage[inline]{trackchanges} 
\usepackage{soul}

\draftfalse

\journalname{arXiv}

\begin{document}
\title{Rediscover Climate Change during Global Warming Slowdown via Wasserstein Stability Analysis}
\authors{Zhiang Xie\affil{1},  Dongwei Chen\affil{2},  and Puxi Li\affil{3}}

\affiliation{1}{Department of Earth and Space Sciences, Southern University of Science and Technology, Shenzhen 518055, Guangdong, China}
\affiliation{2}{School of Mathematical and Statistical Sciences, Clemson University,
Clemson 29641, SC, USA}
\affiliation{3}{State Key Laboratory of Severe Weather, Chinese Academy of Meteorological Sciences, China Meteorological Administration, Beijing 100081, China}

\correspondingauthor{Dongwei Chen}{dongwec@g.clemson.edu}





\justifying

\section*{Abstract}
Climate change is one of the key topics in climate science. However, previous research has predominantly concentrated on changes in mean values, and few research examines changes in Probability Distribution Function (PDF). In this study, a novel method called Wasserstein Stability Analysis (WSA) is developed to identify PDF changes, especially the extreme event shift and non-linear physical value constraint variation in climate change. WSA is applied to 21st-century warming slowdown period and is compared with  traditional mean-value trend analysis. The result indicates that despite no significant trend,  the central-eastern Pacific experienced a decline in hot extremes and an increase in cold extremes, indicating a La Nina-like temperature shift. Further analysis at two Arctic locations suggests sea ice severely restricts the hot extremes of surface air temperature. This impact is diminishing as sea ice melts. Overall, based on detecting PDF changes, WSA is a useful method for re-discovering climate change.

\section*{Keywords}
climate change; probability distribution function; extremes; Wasserstein distance;



\section{Introduction}

How climate system evolves with time is one of the essential topics in climate science. 
The Intergovernmental Panel on Climate Change Sixth Assessment Report (IPCC AR6) pointed out that the global mean of surface air temperature (SAT) follows a general warming trend since 20th century, which is mainly attributed to the human influence. 
Consequently, considerable work is devoted to discussing the mean temperature warming trend and its spatial distribution \cite{Delworth2000,Delworth2015,Huang2017,Li2015}. 

The mean value is one of the most significant statistical characteristics of the data set, as it represents the overall energy condition and even the low-frequency variability of the climate system. 
However, the human-induced mean value warming of SAT may be partially countered by internal variability or external forcing, thus the warming trend is not always strong and rising at the same pace.
As a result, the mean value change, especially the trend based on mean value, cannot give sufficient information about climate change.  
In the first decade of the 21st century, global warming decelerates significantly \cite{Modak2021,Lee2015,Delworth2015,England2014,Li2015}. 
However, during the period, a series of climate change are still recorded, such as extreme events \cite{Johnson2018} and La-Nina-like Pacific temperature pattern \cite{Kosaka2013}.
Unfortunately, it is difficult to detect these climate change signals by trend or mean value analysis since the relevant signals during the warming slowdown period are quite weak.  Recently, a new mathematical tool from optimal transport, called Wasserstein distance (W-distance), sheds more light on this question. 

W-distance $\mathcal{W}_p(\mu,\nu)$ on $\mathbb{R}^d$ is induced by the study of how to transport mass from a probability distribution $\mu$ to another distribution $\nu$ in the cheapest way, which is given by
\begin{equation}\label{eq3}
\mathcal{W}_p(\mu,\nu) = \left(\underset{\pi \in \Pi(\mu,\nu)}{\text{inf}} \int_{\mathbb{R}^d \times \mathbb{R}^d} \left|x-y\right|^p \ d\pi(x,y) \right) ^ {\frac{1}{p}}
\end{equation}
where $p \in [1,\infty)$ and $\pi \in \Pi(\mu,\nu)$  means $\pi$ is a joint distribution of $\mu$ and $\nu$. W-distance satisfies the metric axioms and can quantify the distance and similarities between two  probability distributions \cite{figalli2021invitation}. More details about W-distance could be found in supporting information materials.  In this work, we use earth mover's distance, i.e. $p=1$. 

Wasserstein distance has been applied to climate science, such as model evaluation \cite{vissio2020evaluating}, oceanographic data analysis \cite{hyun2022ocean}, and data assimilation \cite{tamang2022advancing}. However, it is rarely applied in the physical analysis of climate data. Based on W-distance, we develop a novel method, named as Wasserstein Stability Analysis (WSA), to discover PDF variability in climate change, which includes signals such as extremes and physical value constraints. 
Compared to mean value analysis, WSA could help researchers directly identify the significant PDF variability and gives a better insight into climate change.

The paper is organized as follows. In section 2, we give a description to the data set and WSA algorithm. Optimal mass transportation and Wasserstein distance are also briefly introduced. In section 3, we apply WSA to global warming slowdown period. Through the new method, we obtain a clear equatorial eastern Pacific signal that is highlighted in previous studies and an evident physical value constraint change in the Arctic that is not fully explored. Finally, in section 4, we draw a conclusion and discuss the future work.

\section{Data and Method}
\subsection{Dataset}
The surface 2 meters air temperature data in this work are obtained from ERA5 dataset ($0.25^o \times 0.25^o$, single layer, half day time interval) from 1998 to 2012 \cite{hersbach2020era5}. And the spatial range is global. 
To better represent the large-scale signal and  accelerate the data processing, we regrid the original data set into daily data with a horizontal resolution of $2.5^o \times 2.5^o$ by spatial and temporal average. 

The sea ice data used in the study come from Hadley Centre Sea Ice and Sea Surface Temperature monthly data set (HadISST), with 1$^o$ $\times$ 1$^o$ horizontal resolution  \cite{Rayner2003}. The time range is from 1998 to 2012 and the spatial range is still global. 
\subsection{Wasserstein Stability Analysis}


In the null hypothesis of most statistical analyses, we usually assume that the PDF of one variable is stable and follows the same distribution. 
Following this logic, we claim that the variable has an unstable PDF if the W-distance between its current state and a reference state is out of a specific confidence interval, representing a significant change from one climate stage to another. 
Therefore, the significant PDF change signal is detected by testing W-distance, and the climate change details can be further explored. 
Hence, in this part, we develop a novel method  named as Wasserstein Stability Analysis (WSA) to evaluate the magnitude of PDF change quantitatively, i.e., to detect unstable PDF change signals.

The new method is mainly divided into two steps: W-distance test and PDF analysis. For more information about W-distance test, interested readers could refer to Algorithm 1 in the supporting information material. The details are as follows:

(1). \textit{W-distance test}.  Given two samples $X$ and $Y$, both series are first normalized individually by the following scalar
\begin{equation}
    \frac{x-\mu}{\sigma}
\end{equation}
where $\mu$ is the mean value and $\sigma$ is the standard deviation. 
Then Wasserstein distance $\mathcal{W}_1(X,Y)$ between $X$ and $Y$ are obtained after normalized series. Note that normalization is necessary since the geographic difference may lead to a large variation in W-distance. 
For example, the high surface temperature variation at high latitudes usually leads to a larger W-distance than its counterpart in tropical zones. Nevertheless, it does not necessarily mean the PDF change at high latitudes is stronger than that at low latitudes. 

Our significance test is based on the Monte Carro test \cite{Xie2017}. The null hypothesis $H_0$ of WSA is that the PDF discrepancy between $X$ and $Y$ can be explained by the white noise. We generate two new samples, $X_N$ and $Y_N$, by adding white noise to $X$ and $Y$. Then one could obtain a new W-distance $\mathcal{W}_1(X_N,Y_N)$.
Such operation is performed 500 times. Then one could get a confidence interval (C.I.) of W-distances. 
Under the null hypothesis with confidence level $1-\alpha$, $\mathcal{W}_1(X,Y)$ must be indistinguishable from $\mathcal{W}_1(X_N,Y_N)$. Thus one would expect $\mathcal{W}_1(X,Y)$ is within the confidence interval of $\{\mathcal{W}_1(X_N,Y_N)\}_{N=1}^{500}$. If not, $H_0$ is rejected with the confidence level of $1-\alpha$. The $\alpha$ in this study has been set as 0.01 and we perform the W-distance test for all locations for spatial pattern.

(2). \textit{PDF analysis}. After detecting the significant PDF change zones where the W-distance is significant, we plot two PDFs and the corresponding change for each significant change zone, and then explore  the climate change mechanism that makes W-distance significant. 
The next section will use two examples to demonstrate how it functions.

\section{Results}
During 1998 - 2012 period, global mean surface temperature experienced a slower increase, which is referred to as "global warming hiatus" in early studies \cite{Kosaka2013}. 
Recent research indicated that there was no true "hiatus" during that time, but the global surface temperatures were still rising, albeit more slowly \cite{Simmons2017}. 
This phenomenon suggests that physical processes other than the anthropogenic warming trend play a role in climate change, and the detection of PDF change may provide us with new information for this period.  Therefore, by comparing the performance of the WSA with traditional trend analysis, we can evaluate the similarity and difference between these two approaches, which could advance our understanding of climate science and inform future research in this area.

\begin{figure}
    \centering
    \includegraphics[scale=0.6]{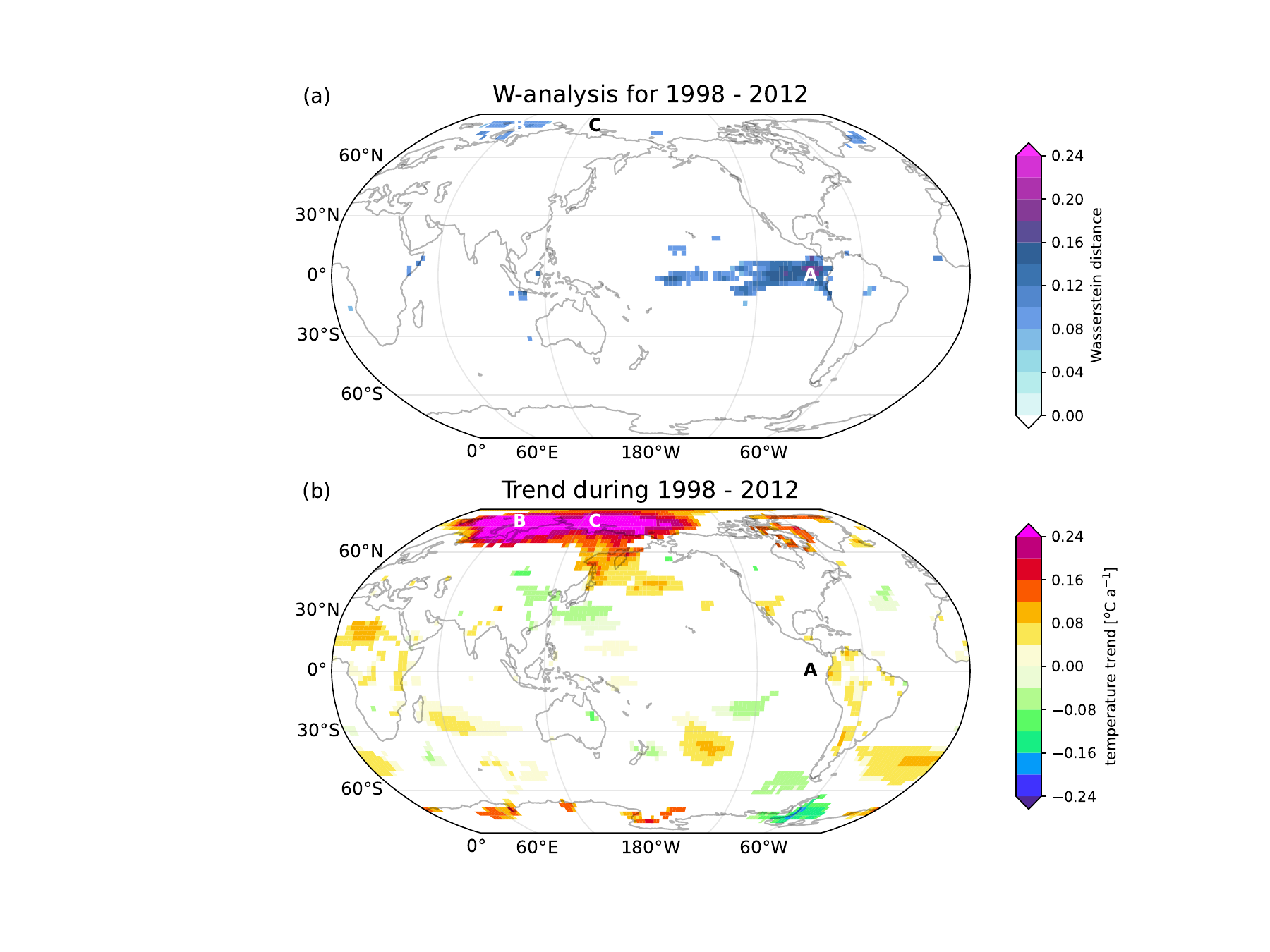}
    \caption{ (a) Wasserstein distance between probability distribution functions during 1998-2005 and 2005-2012. (b) Linear trend during 1998-2012. Only statistically significant regions (99\% for Wasserstein distance and 95\% for linear trend) are shown. Three sites, A (90$^o$W, 0), B (66$^o$E, 79$^o$N), and C (131$^o$E, 79$^o$N), have been chosen for further analysis. }
    \label{fig:wdis_trend}
\end{figure}

Figure \ref{fig:wdis_trend}a depicts the Wasserstein distance between SAT PDFs during 1998-2005 and 2005-2012. 
There are two main large-scale significant W-distance signals appearing in Figure \ref{fig:wdis_trend}a: the central eastern Pacific and the Barents-Kara Sea. 
The central eastern Pacific has the greatest W-distance, with values ranging from 0.08 to 0.24; nevertheless, there is no discernible trend (Figure \ref{fig:wdis_trend}b). 
Another noticeable W-distance signal could be seen over the Barents-Kara Sea, which is located inside a significant warming trend zone in the Arctic (Figure \ref{fig:wdis_trend}b). The following analysis will mainly focus on signals over central eastern Pacific and Barents-Kara Sea to highlight the information that W-distance can provide about climate change. 
In practice, we found that due to the presence of sea ice, the surface temperature distribution shows a large discontinuous in the Arctic region.
As a result, the areal average in this region may smash some information from individual grids. 
To avoid this, we selected three special sites to do further PDF analysis: \\
1. Site A (90 $^o$W, 0): one site with a significant W-distance signal but without a significant linear trend; \\
2. Sites B (66 $^o$E,79 $^o$N) and C (131 $^o$E,79 $^o$N): both sites exhibit significant linear trends. However, Site B has a significant W-distance while C does not.

\begin{figure}
    \centering
    \includegraphics[scale=0.65]{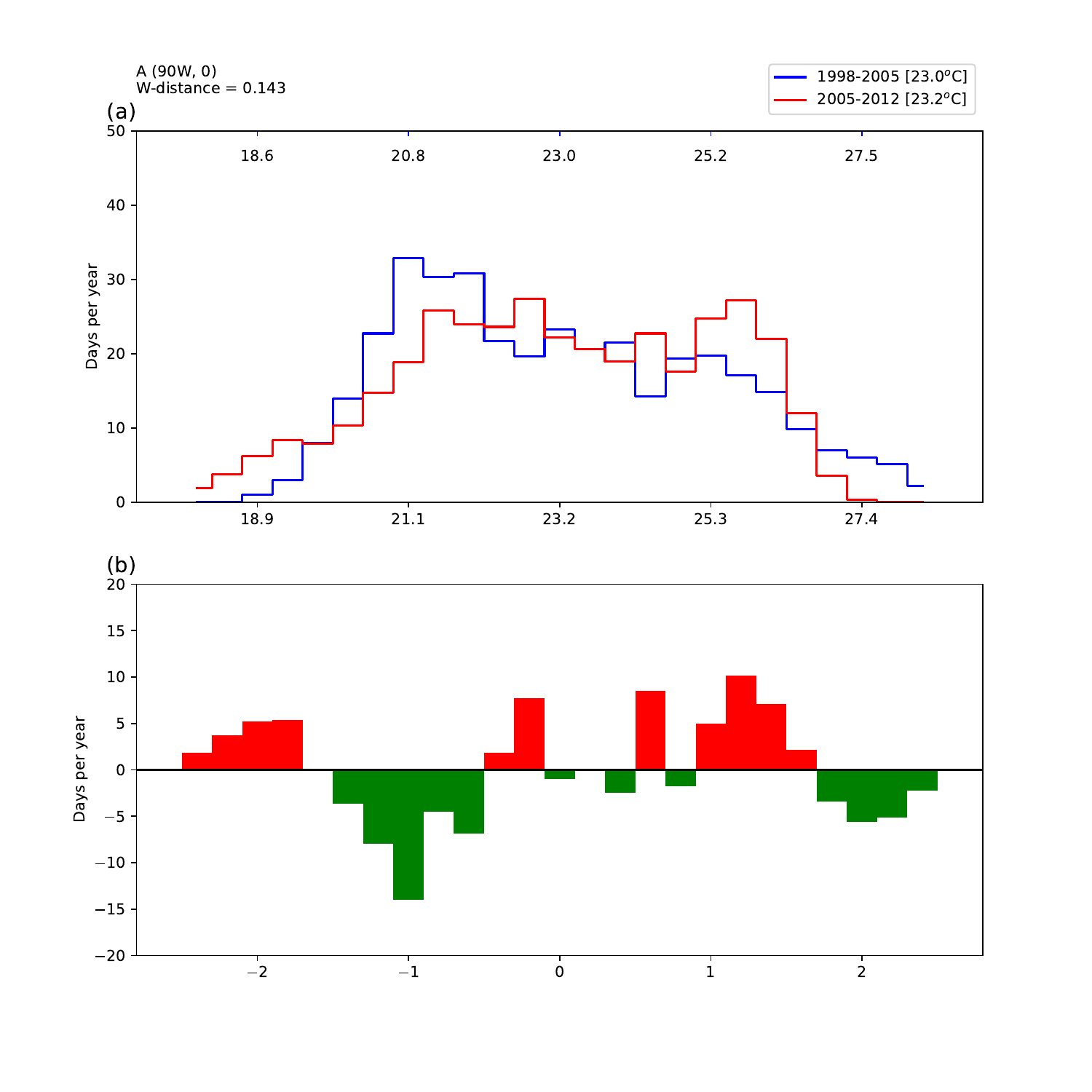}
    \caption{
   (a) Probability distribution functions (PDFs) for the periods 1998–2005 (blue) and 2005–2012 (red).  The bottom tick labels represent observed temperature values from 2005 to 2012, whereas the top tick labels represent observed temperature values from 1998 to 2005. (b) The difference between two PDFs of  normalized temperature series at site A (90$^o$W, 0). The frequency has been normalized as days per year. The mean value for the time period is shown by the blanketed number in the legend.
    }
    \label{fig:siteA}
\end{figure}

Figure \ref{fig:siteA} shows the SAT PDFs and their changes from the 1998-2005 period to the 2005-2012 period at Site A. 
The mean value and standard deviation stay relatively stable within the 1998-2005 and 2005-2012 periods. 
However, in the second period, the hot extremes (more than two standard deviations, roughly 27 $^o$C) decrease by more than ten days per year, while the mild hot events (near one standard deviation) increase by roughly 30 days per year. 
Meanwhile, the cold extremes (less than -2 standard deviations, 19 $^o$C) increase by more than ten days per year, while the mild cold events (near -1 standard deviation) decrease by more than 30 days per year. 
The hot extreme decreasing and cold extreme increasing indicate a La Nina-like SAT shift. 
Many early studies argue that the 21st slowdown period is in the negative phase of Pacific Decadal Oscillation, and this strong low-frequency variability canceled out some parts of the global warming trend
\cite{Douville2015,Kosaka2013,Lee2015,England2014}. 
Our W-distance result agrees well with this conclusion and extends the results from the perspective of PDF change. 
Indeed, even though the mean value does not show a significant trend, we can still observe the La Nina-like SAT shift from extreme event change, which is reflected by the strong PDF change and the corresponding significant W-distance.

\begin{figure}
   \centering
    \includegraphics[scale=0.65]{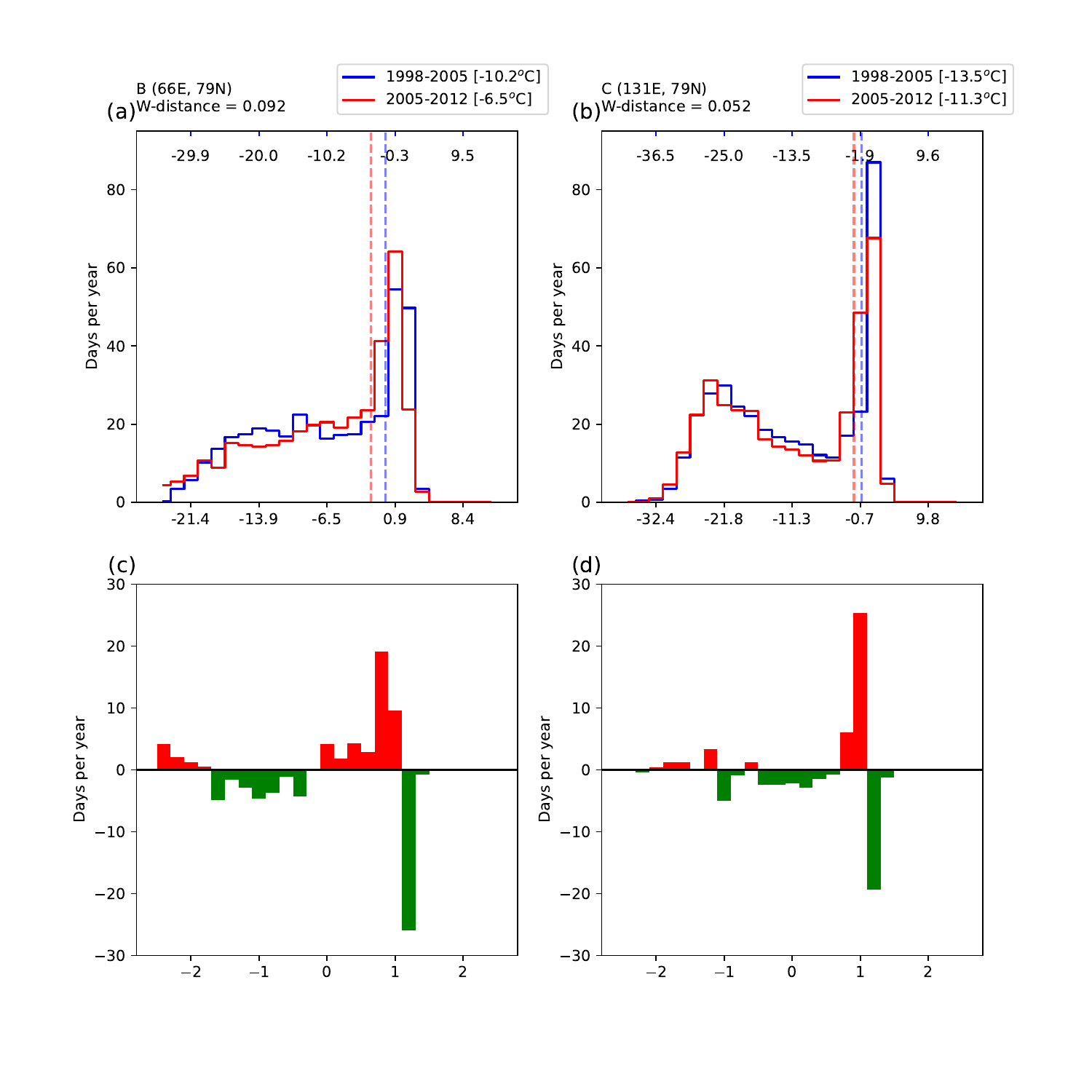}
    \caption{Same as Figure \ref{fig:siteA}. Panel (a) and (c) is for Site B, and Panel (b) and (d) is for Site C. The dashed lines in Panel (a) and (b) represent the ocean frozen temperature (-1.7$^o$C) in two distributions (blue for 1998-2005 and red for 2005-2012). 
    }
    \label{fig:siteBC}
\end{figure}

Site B is another region with a significant W-distance, suggesting a strong PDF change from the 1998-2005 period to the 2005-2012 period (Figure \ref{fig:siteBC} (a) and (c)). 
At Site B, SAT shows significant warming from -10.2 $^o$C to -6.5 $^o$C. 
Meanwhile, the standard deviation also decreases, indicating an elimination of SAT variation. 
Since Site B is located within a zone covered by sea ice, the maximum SAT is strongly constrained to being near the frozen line in both periods, and thus this causes a sharp frequency jump near the frozen line (Figure \ref{fig:siteBC} (a)). 
However, as the mean value increases and the standard deviation decreases during the second period, the frozen line at Site B shifts significantly to the left.
As a result, the frequency peak of hot events shifts to the left. 
Meanwhile, the slope near the frozen line becomes flatter, leading to a smoother PDF shape. 
In contrast, Site C exhibits a change that follows a similar pattern but is considerably smaller in extent in terms of mean value, standard deviation, and frozen line shift (Figure \ref{fig:siteBC} (b) and (d)). 
Consequently, W-distance at Site C is smaller than that at Site B and fails the 99\% significant test. 
Actually, the existence of sea ice strongly limits the maximum surface temperature and the corresponding surface air temperature near the frozen line. 
Therefore, the sharp peak shape of SAT PDF appears in regions covered by sea ice. 
Nevertheless, regions with normal ocean conditions are expected to have a more Gaussian distribution.
Therefore, the alteration in Site B and Site C suggests that the variability in sea ice concentration may play a role in the SAT PDF change from the first to the second period.

\begin{figure}
    \centering
    \includegraphics[scale=0.6]{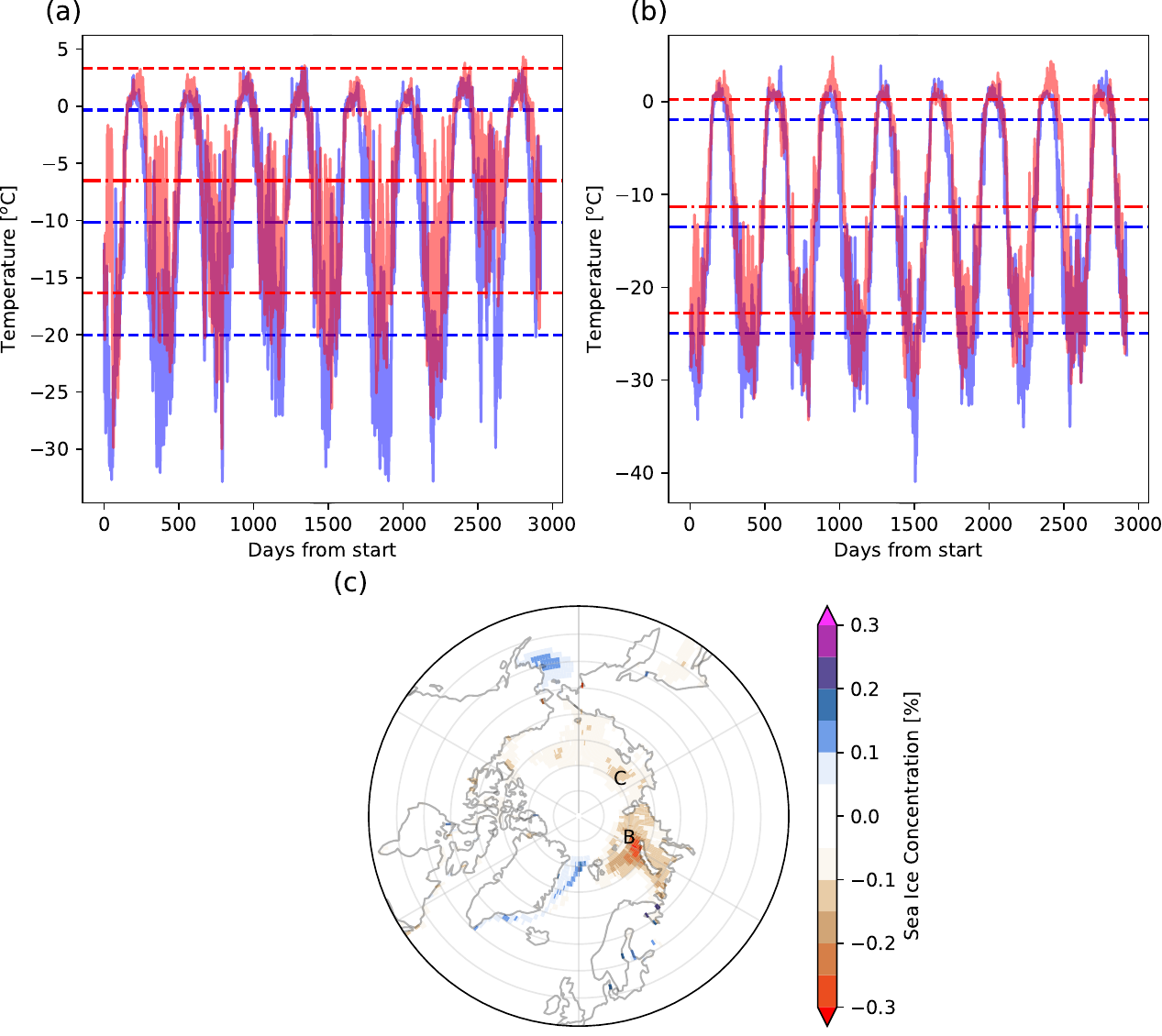}
    \caption{Full daily 2m temperature time series of 1998-2005 (blue) and 2005-2012 (red) at Site B is plotted in Panel (a), and the counterpart at Site C is plotted in Panel (b). Panel (c) shows the sea ice concentration change (unit: \%) between 1998-2005 and 2005-2012.  
    }
    \label{fig:seaice}
\end{figure}

By checking the full-time series in two periods, we can better understand the PDF change at Site B and Site C. 
From 1998-2005 period to 2005-2012 period, the interval with $\pm1$ standard deviation  shrinks and moves upward at Site B because of the change in mean value and standard deviation  (Figure \ref{fig:seaice}(a)). 
Clearly, this change leads to most of the SAT variabilities being below +1 standard deviation and also more cold extremes. 
For Site C, we can see a similar change but in a much smaller magnitude (Figure \ref{fig:seaice} (b)).  
As discussed, this difference can be attributed to the sea ice and its strong temperature constraint. 
Figure \ref{fig:seaice} (c) shows the sea ice concentration change between 1998-2005 and 2005-2012. 
We can see that Site B is in a zone where sea ice is melting the most. Hence the PDF change at Site B is the most intense. 
As a comparison, Site C also experiences a sea ice loss but to a relatively small extent. 
As the temperature increases, the sea ice concentration decreases, and the effect of sea ice restriction gradually fades.  
This is why Site C also displays a similar PDF change but with a less significant level. 
In summary, the significant W-distance change at Site B implies the removal of physical temperature constraint imposed by sea ice that causes the sharp peak near the frozen line in PDF. 
Actually, the sea ice-induced PDF change we discuss here is also related to modelling bias. IPCC AR6 indicated that observational-based estimates of global mean surface temperature face a significant issue in areas where sea ice melts or grows, resulting in a switch between air temperature and sea surface temperature. This bias towards reduced warming in anomalies primarily affects analyses combining surface air temperature anomalies over land and ice with sea surface temperature anomalies over ocean. According to \citeA{Richardson2018}, this underestimated warming in historical model simulations amounts to about 3\% of the observed warming. Given that WSA is able to detect this sea ice-induced change in PDF, it may be useful for future model development and data correction.  

Overall, by detecting changes in the probability density function (PDF) between the early and later periods during slowdown period, WSA is able to identify the two physical processes during the warming slowdown period: the La Nina-like tropical Pacific temperature shift and physical value constraint due to sea ice.

\section{Conclusion and Discussion}
In this work, we develop a novel method called WSA to detect climate change signals. 
The results indicate that WSA shows some advantages compared with the classic mean value analysis in certain cases. The classic trend analysis based on mean value cannot detect the extreme event signals in the central eastern Pacific Ocean during warming hiatus, while WSA can easily identify the La Nina-like temperature PDF shift.  Additionally, due to the strong temperature constraint of sea ice, the sea ice cover area shows a different PDF pattern compared to the open ocean. 
As a result, unlike mean value analysis detecting significant trend regions, WSA can find out the strong sea-ice-loss area during warming hiatus, indicating a fundamental physical property change of the surface. 

It should be noted that the discussion of PDF change during the global warming slowdown period is incomplete in this study. 
We just chose several typical sites to illustrate the abundant information from the significant PDF change and  verify the capability of the WSA algorithm. 
A more detailed discussion on scattered significant signals in the Indian Ocean and South America is another interesting topic. 
Meanwhile, we only applied the WSA in SAT in this study. Similarly, we can also apply the WSA to precipitation, wind speed, and pressure, which may also offer more insights into climate change.

However, the current WSA method could only detect whether the distribution has changed or not, and the direction of change is not well depicted. For example, Fig. \ref{fig:wdis_trend}(a) can only give us a clue to discover PDF change signals, but cannot directly tell us whether most of the regions have more cold extremes or warm extremes. 
Therefore, how to measure the direction of PDF change is still a challenging topic and needs to be explored in the future. 

\section*{Data and Software}
 Surface 2 meters air temperature data is from ERA5 reanalysis products of the European Centre for Medium-Range Weather Forecast (ECMWF) \cite{hersbach2020era5}, which can be accessed  at the following link 
\url{ https://doi.org/10.24381/cds.adbb2d47}. Sea ice data is from the Hadley Centre Sea Ice and Sea Surface Temperature dataset (HadISST) of Met Office in British \cite{Rayner2003}, which could be accessed  via the following link \url{http://www. metoffice.gov.uk/hadobs/hadisst/}. The code of Wasserstein Stability Analysis can be found online via the  link 
\url{ https://doi.org/10.5281/zenodo.7839648}

\section*{Acknowledgement}
The authors greatly appreciate the comments of Dr. Zhaolu Hou from the Ocean University of China and Dr. Shuailei Yao from the Institute of Atmospheric Physics, Chinese Academy of Sciences. This work is supported by the National Natural Science Foundation of China 42005039. 

\section*{Supplementary Material}
This material is the supporting information for Wasserstein distance in optimal transport and  \textit{W-distance test} in Wasserstein Stability Analysis. In the first section, We give an introduction to Wasserstein distance, and especially we list the open source code to compute Wasserstein distance. Then in the second section, we give a detailed description of the \textit{W-distance test} of Wasserstein Stability Analysis, which is formulated in Algorithm \ref{alg:cap}

\section*{1. Introduction to Wasserstein Distance and Optimal Transport}

This section gives an introduction to optimal transport and Wasserstein distance(W-distance). In practice, W-distance can quantify the distance and similarities between two  probability distributions. Compared to other similarity metrics like Kullback–Leibler divergence, Wasserstein distance is rigorously defined and satisfies the metric axioms \cite{figalli2021invitation}. Wasserstein distance provides a powerful metric to quantify the similarities and discrepancies between probability distributions, and has been applied to climate science, such as model evaluation \cite{vissio2020evaluating}, oceanographic data analysis \cite{hyun2022ocean}, and data assimilation \cite{tamang2022advancing}. 

In 1781, Gaspard Monge proposed the concept of optimal transport from one practical situation: if one uses soil to build fortifications, what is the cheapest way to transport the soil?  Let $\mu$ and $\nu$ be two probability measures (distributions) on $\mathbb{R}^d$, this scenario leads to the following Monge formulation of optimal transport
\begin{equation}\label{eq1}
 \underset{ T_{\#} \mu = \nu}{\text{inf}} \ \int_{\mathbb{R}^d} c(x, T(x)) \ d\mu(x)
\end{equation}
where  $c(x,T(x)) $ is the cost  of transporting unit mass from $x$ to $T(x)$, and $\nu = T_{\#} \mu$ means that for any (measurable) set $A \subset \mathbb{R}^d$,
\begin{equation}
    \nu(A) = \mu(T^{-1}(A))
\end{equation}
Such $T$  is said to be a transport map, and since  $T$ is a map, the mass at $x$ could only be transported to one destination, which means Monge formulation does not allow splitting mass.  

In the 1940s, Leonid Kantorovich revisited Monge's problem and gave relaxation to Monge's formulation by splitting the mass. Kantorovich relaxation leads to the following formulation of optimal transport
\begin{equation}\label{eq2}
 \underset{\pi \in \Pi(\mu,\nu)}{\text{inf}}  \int_{\mathbb{R}^d \times \mathbb{R}^d} c(x,y) \ d\pi(x,y)
\end{equation}
where $\pi \in \Pi(\mu,\nu)$  means $\pi$ is a joint distribution  with marginals $\mu$ and $\nu$, i.e.
\begin{equation}
    \Pi(\mu,\nu)=\{\pi \in \mathcal{P}(\mathbb{R}^d \times \mathbb{R}^d): \pi(A \times \mathbb{R}^d)= \mu(A), \pi(\mathbb{R}^d \times B)= \nu(B), A, B \subset \mathbb{R}^d \}
\end{equation}

When the cost function $c(x,y) = \left|x-y\right|^p$, the metric side of Kantorovich formulation makes it valid to quantify the similarities between probability measures $\mu$ and $\nu$ via p-Wasserstein distance, defined as below
\begin{equation}
\mathcal{W}_p(\mu,\nu) = \left(\underset{\pi \in \Pi(\mu,\nu)}{\text{inf}} \int_{\mathbb{R}^d \times \mathbb{R}^d} \left|x-y\right|^p \ d\pi(x,y) \right) ^ {\frac{1}{p}}
\end{equation}
where $p \in [1, \infty)$.  In this work, we use the following earth mover's distance, i.e. $p=1$,

\begin{equation}
\mathcal{W}_1(\mu,\nu) = \underset{\pi \in \Pi(\mu,\nu)}{\text{inf}} \int_{\mathbb{R}^d \times \mathbb{R}^d} \left|x-y\right| \ d\pi(x,y) 
\end{equation}

Earth mover’s distance can be seen as the minimum amount of “work” required to transform mass from the probability measure (distribution) $\mu$ into another probability measure (distribution) $\nu$, where the “work” is measured as the amount of distribution weight that must be moved, multiplied by the distance it has to be moved.

The code to calculate the earth mover's distance between two $1D$ distributions has been integrated into the $wasserstein\_distance$ function of \textit{Scipy} package. The input of $wasserstein\_distance$ is just the
two sequences of observed values in the empirical distributions, then the function returns the computed distance between these two distributions. The details can be found at \cite{wdistance2023earthMover}. 

Another option to compute Wasserstein distance is to use the $ot.emd2$ function in Python optimal transport package $POT$ \cite{flamary2021pot}. The $ot.emd2$ function could compute the earth mover distance in $1D$ as long as the cost matrix is given by the corresponding $l_1$ distance. While $POT$ is one of the most efficient exact optimal transport solvers, it has not been designed to handle large-scale optimal transport problems. Therefore, if one needs to solve optimal transport with large number of samples, we do not recommend to use  $POT$. Interested readers may  refer to $POT$ homepage \url{https://PythonOT.github.io/} for more information.

\section*{2. W-distance Test in Wasserstein Stability Analysis}

Algorithm \ref{alg:cap} provides details for the first step \textit{W-distance test} in Wasserstein Stability Analysis.  Given two samples $X$ and $Y$, both series are first normalized individually by the following scalar
\begin{equation}
    \frac{x-\mu}{\sigma}
\end{equation}
where $\mu$ is the mean value and $\sigma$ is the standard deviation. 
Then Wasserstein distance $\mathcal{W}_1(X,Y)$ between $X$ and $Y$ are obtained after normalization.

It is worth noting that this normalization is necessary since the geographic difference may lead to a large variation in W-distance. 
For example, the high surface temperature variation at high latitudes usually leads to a larger W-distance than its counterpart in tropical zones. Nevertheless, it does not necessarily mean the PDF change at high latitudes is stronger than that at low latitudes. Furthermore, the length of the two samples $X$ and $Y$ in WAS could be different.

Our significance test is based on the Monte Carro test \cite{Xie2017}. The null hypothesis $H_0$ of WSA is that the PDF discrepancy between two samples $X$ and $Y$ can be explained by the white noise. Clearly, the alternate hypothesis ($H_1$) is that the discrepancy of PDFs between $X$ and $Y$ results from factors other than white noise.

After getting Wasserstein distance $\mathcal{W}_1(X,Y)$, we generate two new samples, $X_N$ and $Y_N$, by adding white noise $N(0,1)$ to $X$ and $Y$, where $N(0,1)$ is the standard Gaussian distribution.
Then one could obtain the W-distance $\mathcal{W}_1(X_N,Y_N)$ between the new samples $X_N$ and $Y_N$. 
Such additional operation of white noise is performed 500 times. Therefore, one could get a sequence of W-distances $\{\mathcal{W}_1(X_N,Y_N)\}_{N=1}^{500}$, and then get a confidence interval (C.I.) of W-distances. 
Under the null hypothesis $H_0$ that the distribution discrepancy between $X$ and $Y$ can be explained by the white noise $N(0,1)$, the W-distance between the original samples $X$ and $Y$ must be indistinguishable from the counterpart distance between perturbed samples  $X_N$ and $Y_N$, i.e.,  $\mathcal{W}_1(X,Y)$ must be indistinguishable from $\mathcal{W}_1(X_N,Y_N)$.
Under the null hypothesis $H_0$ with the confidence level $1-\alpha$, one would expect that $\mathcal{W}_1(X,Y)$ is within the confidence interval of $\{\mathcal{W}_1(X_N,Y_N)\}_{N=1}^{500}$. If not, $H_0$ is rejected, and one could claim that the distribution discrepancy between $X$ and $Y$ is not from the white noise $N(0,1)$ with the confidence level of $1-\alpha$, where $\alpha=0.01$.

In this study, a higher confidence level of $1-\alpha =  99\%$, rather than the common $ 95\%$, is used, since W-distance is very sensitive to the PDF change. Furthermore, when adding white noise to the normalized samples of $X$ and $Y$, we use the standard Gaussian distribution $N(0,1)$. Since a large size of samples is used, according to the central limit theorem, the amplitude of white noise is set as one standard deviation individually. 

After performing the W-distance test for all locations, we could detect the places where the W-distance is significant, which are regarded as the significant PDF change zones. After getting the significant PDF change zones, we could do \textit{PDF analysis}: plot two PDFs and the corresponding change for each significant change zone, to understand the climate change mechanism which makes W-distance significant.

\begin{algorithm}
\caption{Wasserstein Stability Analysis (WSA)}\label{alg:cap}
\begin{algorithmic}[1]
\Procedure{}{$H_0$: the discrepancy in distributions between $X$ and $Y$ is from white noise, with confidence level $\alpha = 0.01$}
\State $X$ and $Y$ are normalized individually by the $\frac{x-\mu}{\sigma}$ scalar.
\State Get $\mathcal{W}_1(X,Y)$
\State $N \gets 1$
\While{$N \leq 500$} 
    \State $X_N \gets X+random(0,1)$ \Comment{add white noise $N(0,1)$ to $X$ and  $Y$ }
    \State $Y_N \gets Y+random(0,1)$
\State Get $\mathcal{W}_1(X_N,Y_N)$ \Comment{$X_N$ and $Y_N$ are series with white noise}
\State $N \gets N+1$
\EndWhile
\If{$\mathcal{W}_1(X,Y)$ is out of $99\%$ C.I. of $\{\mathcal{W}_1(X_N,Y_N)\}_{N=1}^{500}$}
    \State reject $H_0$ \Comment{  The discrepancy is not  from $N(0,1)$}
\Else 
\State $H_0$ is not rejected
\EndIf
\EndProcedure
\end{algorithmic}\label{algo:WSA}
\end{algorithm}

\newpage
\bibliography{My_citation}

\end{document}